\documentclass{ltjournalarticle}
\usepackage[dvipsnames]{xcolor}
\usepackage{lineno,hyperref,xcolor,float,caption,paralist,siunitx,amssymb} 
\usepackage[singlespacing]{setspace}
\usepackage{soul} 
\modulolinenumbers[5]
\usepackage[T1]{fontenc}
\usepackage{wrapfig}
\usepackage{multicol}
\usepackage{changes}

\usepackage{amssymb}
\usepackage{amsthm}
\usepackage{graphicx} 
\usepackage{amsmath}
\usepackage{upgreek}
\usepackage{float}
\usepackage{xcolor}
\usepackage{subcaption}
\usepackage{siunitx}
\usepackage{gensymb}
\usepackage[T1]{fontenc}
\usepackage{placeins}
\usepackage{ulem}
\usepackage[utf8]{inputenc}
\usepackage{amsmath}
\usepackage{graphicx}
\usepackage{float}
\usepackage{appendix}
\usepackage{caption}
\usepackage{subcaption}
\usepackage{siunitx}
\usepackage{xcolor}
\usepackage{tikz}
\usepackage{color}
\usepackage{upgreek}
\usepackage{textgreek}
\usepackage{xspace}
\usepackage{wrapfig}
\usepackage{textcomp}
\usepackage{booktabs}
\usepackage{amssymb,amsthm,amsmath,amstext,amsbsy,amsopn}
\usepackage{isotope}
\usepackage{multirow}

\title{Production of hypernuclei from antiproton capture within a relativistic transport model}

\author{
 A. Schmidt\textsuperscript{a}\thanks{Corresponding author: \href{mailto:aschmidt@ikp.tu-darmstadt.de}{aschmidt@ikp.tu-darmstadt.de}},
 T. Gaitanos\textsuperscript{b}, 
 A. Obertelli\textsuperscript{a}, 
 and J.-L. Rodr\'{i}guez-S\'{a}nchez\textsuperscript{c}
}

\date{\footnotesize\textsuperscript{\textbf{a}}Technische Universität Darmstadt, Fachbereich Physik, Darmstadt 64289, Germany\\ \textsuperscript{\textbf{b}}Physics Department, School of Physics, Aristotle University of Thessaloniki, Thessaloniki 54124, Greece\\ \textsuperscript{\textbf{c}}CITENI, Campus Industrial de Ferrol, Universidade da Coru\~{n}a, Ferrol 15403, Spain}

\renewcommand{\maketitlehookd}{
	\begin{abstract}
		\noindent Production yields of single-\textLambda\,hypernuclei from simulated peripheral annihilations of antiprotons after capture on various target nuclei are reported. The initial annihilation process and the production of excited hypernuclei are estimated within the GiBUU transport framework, while their deexcitation process is treated with the ABLA++ code. The yield of excited hypernuclei range from $\sim$0.3\,\% for \isotope[16][]{O} to $\sim$1.2\,\% for \isotope[132][]{Xe} per annihilation, consistent with previous measurements at LEAR, CERN. The yield of specific ground state hypernuclei after deexcitation reaches up to a few $10^{-4}$ per annihilation. The hypernuclei are produced in strangeness exchange reactions occuring between a nucleon of the target and the kaons originating from the annihilation in $\sim$80\,\% of the cases, while the strangeness pair production in secondary pion-nucleon collision contributes to the remaining $\sim$20\,\%. The simulations indicate that antiproton annihilations at rest on different nuclei could populate a wide range of so-far unexplored hypernuclei. 
	\end{abstract}
}

\begin{document}
\maketitle 

\section{Introduction}
\label{intro}

Knowledge of the baryon-baryon interactions is an essential premise for the ab initio description of baryonic few- and many-body systems. In the nucleonic (N) sector, these interactions can be described by phenomenological boson-exchange models~\cite{MACHLEIDT19871,Machleidt01} or using potentials derived from chiral effective field theory~\cite{Machleidt_2005,hebeler2015nuclear}, relying on precise existing NN-scattering data as well as measured binding energies of light nuclei.\\
The introduction of a hyperon (Y) inside a nucleus additionally requires data to constrain the interaction of Y with nucleons~\cite{gal2016strangeness}. In case of the lightest hyperon, the \textLambda-baryon (valence quarks $u,\,s,\,d$), the \textLambda N interaction has a short attractive range, typically modelled via the exchange of pions or heavier mesons~\cite{haidenbauer2017lambda}. Additional contributions arise from the allowed strange meson exchange as well as the strong \textLambda N-\textSigma N coupling~\cite{akaishi2000coherent,myint2000neutron,Haidenbauer_2021_on}. \\
The latter effect is also essential to reproduce astrophysical observations of neutron stars of two solar \linebreak masses~\cite{antoniadis2013massive,cromartie2020relativistic} with models that account for the equilibrium of production and decay of strange \linebreak baryons in the dense core of neutron stars~\cite{lonardoni2015hyperon}, but additional experimental data are needed to constrain the parameters of the respective models. 

\noindent Historically, experiments focused on the investigation of the scattering cross sections of \textLambda\,baryons on protons, mostly in the final state of a kaon- or pion-induced strangeness production~\cite{Alexander_61,groves1963elastic,Alexander_68,rowley2021improved}. Recently, \textSigma$p$ scattering was investigated at J-PARC~\cite{nanamura2022measurement,Miwa2022precise}. Additional constraints on the \textLambda $p$ and \textLambda $p p$ interaction are obtained by analyzing particle correlations via the femtoscopy technique applied to ultra-relativistic $p$-$p$ and $p$-Pb collisions measured by the ALICE experiment at the LHC~\cite{acharya2019study,acharya2023constraining,acharya2023towards}. In contrast to protons, there are no experimental data on \textLambda $n$ scattering. 
\noindent Properties of the YN interactions inside a nuclear environment can also be extracted from the spectroscopy of bound systems of nucleons and hyperons, {\itshape i.e.}, hypernuclei. Several phenomenological \linebreak models~\cite{Gal71,Gal72,Gal78,Rayet81,Moto83,Rufa87,Mares89,Tret99,Cugon00,Hiya09,Mill08,Mill10,Mill12} and ab initio quantum Monte Carlo and no-core shell models have been made possible for such hypernuclei~\cite{Lona13,Lona14,Wirth14,Wirth16,Wirth18,Wirth18b}.
The experimental study of hypernuclei to constrain the \textLambda N interaction has been performed by investigating their mass~\cite{bouyssy1976hypernuclei}, \textgamma-decay of excited states~\cite{hackenburg1983observation,may1997first,akikawa2002hypernuclear,ukai2004hypernuclear,tamura2005gamma,ukai2008gamma,tamura2013gamma} and mesonic weak decay~\cite{agnello2009new,agnello2012evidence}. For that, a hypernucleus is typically produced by a strangeness exchange (K$^-$,\textpi$^-$) reaction~\cite{faessler1973spectroscopy,bruckner1976strangeness,bertini1981neutron}, a strangeness pair creation such as the (\textpi$^+$,K$^+$) process~\cite{milner1985observation,pile1991study,hasegawa1996spectroscopic} or an electromagnetic ($e,e'\mathrm{K}^{+}$) reaction~\cite{miyoshi2003high,dohrmann2004angular}. \\  
\noindent Pioneering experiments in 1973~\cite{kerman1973superstrange} showed that ion collisions with energies in the order of GeV can also be used for the production of light hypernuclei. Related experiments at Dubna in the 1980s used light projectiles at energies of about 2\,GeV/nucleon and a \isotope[12][]{C} target~\cite{bando1989production}, followed by the HypHI experiment at GSI~\cite{Ekawa_22,saito2023the}. At such energies, the main production mechanism is the elementary NN\,$\rightarrow$\,\textLambda KN reaction with a threshold energy at 1.6\,GeV. Heavier projectiles and targets, {\itshape e. g.} Ag-Ag and Au-Au, are used by the HADES~\cite{Spies_2020} and STAR~\cite{timmins2009overview} collaborations. Ultra-relativistic $p$-Pb and Pb-Pb collisions at the ALICE experiment at LHC also allow the reconstruction of light hypernuclei~\cite{alice_22} emerging from the freeze-out of a quark-gluon plasma~\cite{andronic2018decoding,saito2021new}. \\
\noindent Simulations also indicate the possible population of heavier neutron-rich hypernuclei in the fragmentation of residues with radioactive ion beams~\cite{buyukcizmeci2012mechanisms}. A dedicated experiment based on charge exchange reactions with heavy-ion projectiles has been proposed~\cite{saito2021novel}. \\
\noindent Hypernuclei can also be produced in a strangeness exchange reaction (K$^- p \rightarrow $ \textpi$^0$ \textLambda\,and K$^- n \rightarrow $ \textpi$^-$ \textLambda) following an antiproton annihilation on a nucleus or via a strangeness pair production in a pion-nucleon interaction~\cite{tsushima2000strangeness}. Such hypernuclei have first been experimentally observed via delayed fission following antiproton stopping at the CERN-PS-177 experiment at the low-energy antiproton ring (LEAR) with an estimated production rate of about 0.3\,\% to 0.7\,\% per annihilation~\cite{bocquet1987delayed,armstrong93} on Bi and U targets, respectively. Additionally, the future PANDA experiment at FAIR plans to investigate the production of double-\textLambda\,hypernuclei using a two-step reaction~\cite{gaitanos2012formation,sanchez2014hypernuclear}.   \\
In this article, the production of \textLambda\,hypernuclei using low-energy antiprotons following the formation of antiprotonic atoms is investigated. In antiprotonic atoms, the annihilation in the nuclear density tail occurs at rest~\cite{doser2022antiprotonic}, which is simulated in the present work by an incident collision energy of few eV. Previous simulation studies considered antiproton-induced reactions for momenta in the range of GeV/$c$, thus investigating a different regime of the antiproton-nucleus interaction~\cite{cugnon1990strangeness,feng20}, where elastic and charge-exchange channel become competitive with the annihilation~\cite{carbonell2023comparison}. Via the simulations, the yields of specific hypernuclei with different stable target nuclei are estimated, the production mechanism is discussed and the target's isospin dependence on the yields is investigated. 

\section{Framework}
\label{framework}

\begin{figure}[hptb]
    \centering
    \includegraphics[width=.98\linewidth]{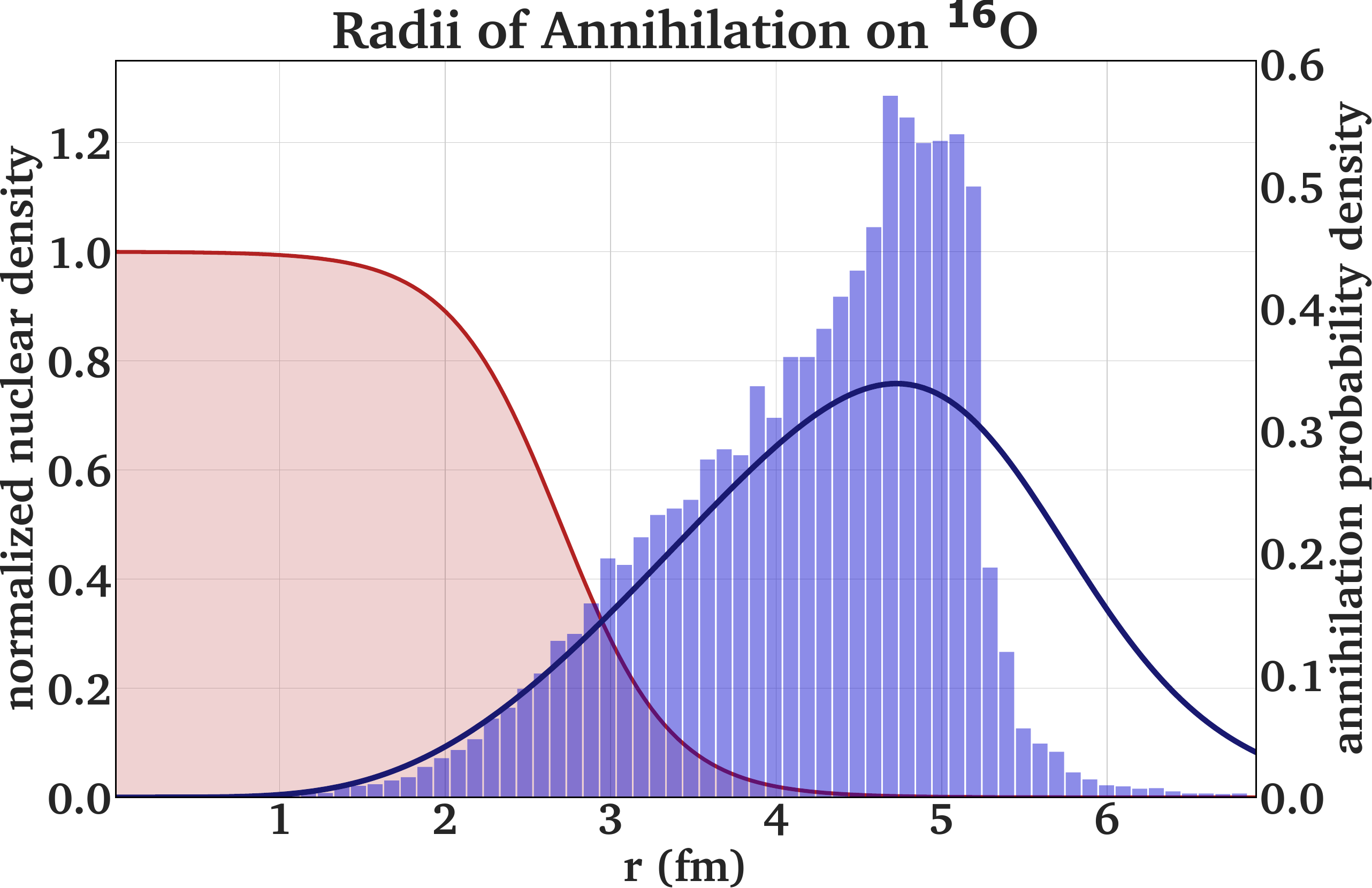}\\
    \vspace{0.2cm}
    \includegraphics[width=.98\linewidth]{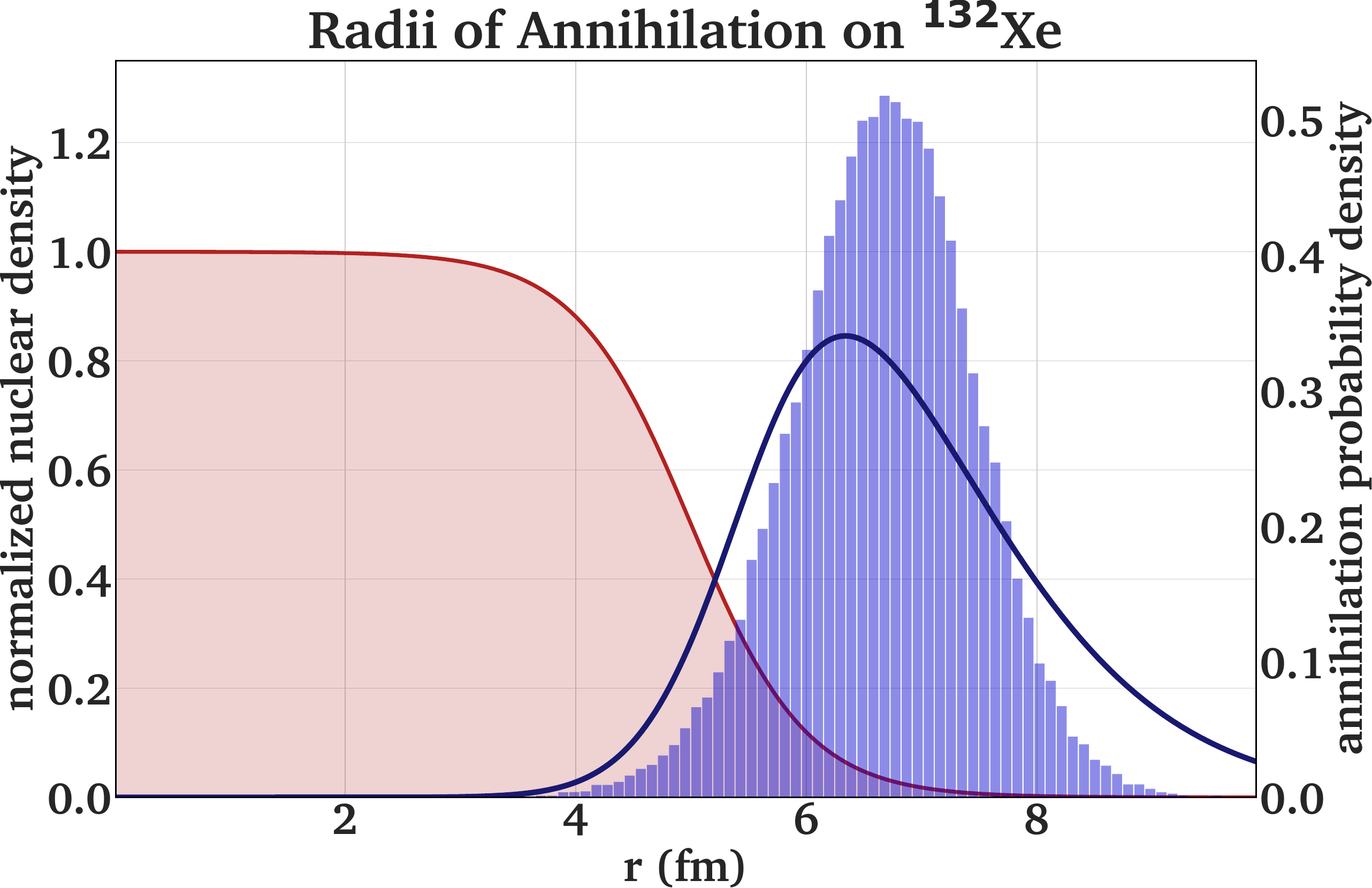}
    \caption{Antiproton annihilation probability density simulated in GiBUU (blue histograms) compared to a prediction based on the antiprotonic radial wavefunction for absorption (solid blue lines) for the examples of \isotope[16][]{O}{} and \isotope[132][]{Xe}{}, where $r$ represents the radial distance to the center of the nucleus. The mean and median values of the distributions are similar, but the particle-based GiBUU simulation underestimates the width compared to the QED-based prediction. The corresponding nuclear density profiles are shown in rose.}
    \label{fig:comparison_rad_an}
\end{figure}

\begin{figure}[hptb]
    \centering
    \includegraphics[width=.85\linewidth]{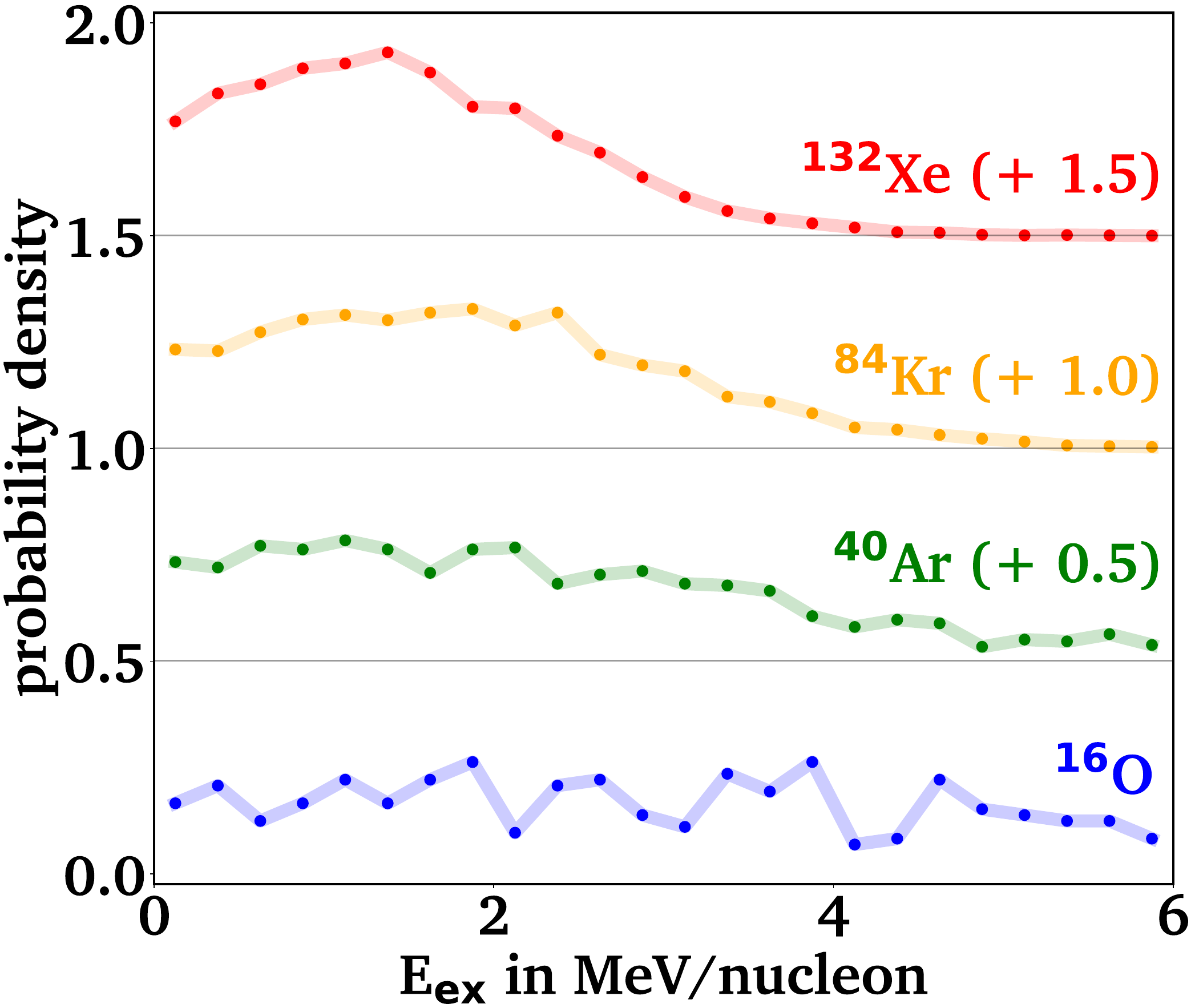}
    \caption{Distribution of the excitation energy E$_\mathrm{ex}$ of the hypersources as simulated by GiBUU. The connecting lines guide the eye and a vertical offset is included for better visibility, as indicated in parentheses.}
    \label{fig:ar_pre_abla}
\end{figure}

Antiprotonic atoms are first formed by the collision of an incident antiproton with a single or multiple bound atomic electrons. After one of these collisions, if the kinetic energy of the antiproton is similar to the electron binding energy, it can be captured by the residual Coulomb potential into a bound antiprotonic orbit~\cite{doser2022antiprotonic}. The principal quantum number of the captured antiproton $n_\mathrm{\bar{p}}$ scales with the principal quantum number of the knocked out electron $n_\mathrm{e}$ and the square-root of quotient of antiproton and electron masses as $n_\mathrm{\bar{p}} \approx \sqrt{m_\mathrm{\bar{p}} / m_\mathrm{e} } \cdot n_\mathrm{e}$. This implies that the antiproton is captured in an orbital with similar size and binding energy $E_\mathrm{B} \sim m_{e ( \bar{p} ) } / n_{e ( \bar{p} )}^2$ as the knocked-out electron. Typically, states with high angular momentum are populated due to the higher number of available substates~\cite{von1987interaction,cohen2000multielectron}. From this highly excited state, the antiproton deexcites via the emission of atomic \linebreak Auger electrons and X-rays, favoring the occupation of circular states ($n$, $l=n-1$) during the cascade~\cite{horvath2013electromagnetic}. For principal quantum numbers of $n_\mathrm{\bar{p}}$\,=\,3\,-\,8, the orbitals start to overlap with the nuclear wavefunction, leading to an annihilation process for small overlap, {\itshape i.e.}, in the tail of the nuclear density~\cite{von1987interaction}. This is consistent with measurement results from LEAR~\cite{pignone94}. The mesons produced in this annihilation at the nuclear surface can then interact with the residual nucleus via final state interactions (FSI), which eventually lead to the production of strange baryons. As the full process, from capture to annihilation, occurs over a wide range of energy scales and interactions, it cannot be treated fully microscopically. This paper focuses on the final annihilation process and presents approximate simulations in the form of ultra-peripheral collisions with low relative energy. 
All simulations presented are performed in two steps. In the first step, the peripheral annihilation of an antiproton with low relative momentum and the subsequent production of an excited \textLambda\,hypernucleus are simulated with the Gießen-Boltzmann-Uehling-Uhlenbeck (GiBUU) \linebreak transport framework~\cite{buss2012transport}. The deexcitation of the excited hypernucleus into a ground state via nucleon, \textLambda \, and photon emission is then performed in the second step with the ABLA++ evaporation and fission module~\cite{kaitaniemi2011incl}.\\ 
The transport code GiBUU and the simulated flow of particles in a reaction are based on the relativistic Boltzmann-Uehling-Uhlenbeck equation~\cite{mori1952quantum}, \linebreak which reads 
\begin{equation}
    \left[k^{* \mu} \partial_\mu^x+\left(k_\nu^* F^{\mu \nu}+m^* \partial_x^\mu m^*\right) \partial_\mu^{k^*}\right] f\left(x, k^*\right)=\mathcal{I}_{\text {coll}}.
\end{equation}
It describes the evolution of the one-body phase space distribution function $f\left(x, k^*\right)$ for hadrons \linebreak within a hadronic relativistic mean field with the kinetic four-momentum $k^* = k - V$, the vector potential $V$ in the form of an in-medium self-energy which depends on the meson currents, the field tensor $F^{\mu \nu} = \partial^\mu V^\nu - \partial^\nu V^\mu$, the effective mass $m^* = m + S$ with the scalar potential $S$ and the contribution of binary collisions and resonance decays $\mathcal{I}_{\text {coll}}$. The hadron propagation in the mean field and scattering cross sections for binary collisions are applicable for a typical energy range of tens of MeV to tens of GeV. As the mesons produced in an annihilation of an antiproton and a nucleon have a typical energy of tens to hundreds of MeV, their propagation is thus suitably described by the framework.\\
So far, GiBUU has been used for simulating numerous phenomena such as the dilepton production at SIS-18/GSI energies~\cite{weil2012dilepton,larionov2020dilepton}, neutrino-induced reactions~\cite{lalakulich2013neutrino}, proton-induced reactions~\cite{gaitanos2008fragment} and the formation of hypernuclei in high-energy reactions~\cite{gaitanos2012formation,gaitanos2009formation,gaitanos2014production}. Owing to the experience in hypernuclear physics and the implementation of antiprotons as projectiles, including realistic branching ratios for the annihilation process, GiBUU was chosen for modeling the production of hyperons based on peripheral antiprotonic annihilations.\\
To prepare an annihilation process at rest, the antiproton is defined as a projectile with a low kinetic energy of 10 eV, while the nucleus of interest is initialized as a target at rest. A variation of the antiproton projectile energy up to 100\,eV did not affect the simulation results. For a given proton and neutron number, the respective density profiles are then derived in the relativistic Thomas-Fermi model~\cite{haddad1993finite}. The impact parameter of the collision depends on the nucleus of interest and is set to reproduce the radial annihilation profile observed in antiprotonic atoms, which typically has a median radius 2\,fm outside the half-density radius of the nucleus. A comparison of the annihilation site distribution derived from the GiBUU simulations with distributions derived by the QED wavefunction overlap for $^{16}$O and $^{132}$Xe as target nuclei are shown in Fig. \ref{fig:comparison_rad_an}. The theoretical distributions are derived from the overlap of the nuclear density with the antiprotonic orbitals, at which the annihilation occurs. In the case of \isotope[16][]{O}{}, a mixing of $n_\mathrm{\bar{p}} = 3,\,4$ (20\,\%, 80\,\%) is assumed based on absorption width measurements in water targets~\cite{kohler1986precision,rohmann1986measurement}, and for \isotope[132][]{Xe}{} a mixing of $n_\mathrm{\bar{p}} = 7,\,8$ (20\,\%, 80\,\%) is interpolated from X-ray intensity measurements on tin isotopes~\cite{schmidt2003nucleon}. It is observed that an annihilation probability density derived from the antiproton radial wavefunctions in the orbitals of annihilations leads to a wider distribution than the simulated annihilation locations in the particle-based transport code GiBUU, leading to an underestimation of super-peripheral annihilation processes. This effect is pronounced strongly for the lighter nuclei, where a sharp drop of the annihilation probability is observed at high radii. However, the mean and median radius of both distributions are similar, so that the majority of annihilation locations are well reproduced.
The annihilation process of the antiprotons with protons and neutrons in GiBUU is modelled with phenomenological branching ratios for the annihilation products, as summarized in appendix A. The mesons produced in the final state of the annihilation can then interact in a second step with the residual nucleons to produce a \textLambda\,baryon via direct strangeness exchange with a primary kaon ({\itshape i.e.}, K$^- n \rightarrow $ \textLambda \textpi$^-$, K$^- n \rightarrow $ \textSigma$^{-,0}$ \textpi$^{0,-}$, K$^- p  \rightarrow $ \textLambda \textpi$^0$ and K$^- p  \rightarrow $ \textSigma$^{+,-,0}$ \textpi$^{-,+,0}$), which is produced in about 5\,\% of annihilations, or via $s$-$\Bar{s}$ pair production induced by a pion ({\itshape i.e.}, \textpi $\mathrm{N} \rightarrow $ \textLambda K \, and \textpi $\mathrm{N} \rightarrow $ \textSigma K for all possible charge combinations). The momentum-dependent cross sections for the individual processes and their implementation in GiBUU are detailed in~\cite{buss2012transport}, and are taken from~\cite{tsushima2000strangeness} for the pion-induced strangeness production and from~\cite{baldini1988total} for the strangeness exchange with kaons. In the next step, the decision if the produced \textLambda\,leads to the formation of a hypersource is taken based on phase space coalescence in the final state of the initial collision, {\itshape i.e.}, at points in time well beyond their production. A hypersource therefore corresponds to an excited hypernucleus after the collision phase and before the statistical decay phase. If the baryon density at the location of the \textLambda\,exceeds the cutoff density of $\rho_\mathrm{c} = 0.01\,\rho_0$, it is attributed to a hypersource, similarly to all other baryons fulfilling the condition. For this hypersource, the total energy is then calculated as the sum over the particles' momenta and the mean field potentials at their locations.\\
To extract the excitation energy of the hypersource, the mass of the respective ground state nucleus has to be subtracted from the determined total energy by assuming a modified Bethe-Weizsäcker mass formula for hypernuclei, which is fitted to experimental data on hypernuclear binding and separation energies~\cite{samanta2006generalized}. This typically leads to excitation energies in the order of 0 to 6 MeV/nucleon, as depicted in Fig. \ref{fig:ar_pre_abla}. 

\begin{table*}[hptb]
\centering
\caption{Simulated statistics of the antiproton annihilations in GiBUU. Hypersources (HS) are excited hypernuclei, which are defined by phase-space coalescence. Only statistical uncertainties are presented. The initial sample for each nucleus includes 250,000 collisions.}
\begin{tabular}{ c  c  c  c  c  c  c }
\toprule

Nucleus & annihilations & \textLambda s & \textLambda\,rate in $\%$ & HSs & HS rate in $\%$ & HS per \textLambda\\

\midrule 

\isotope[16][]O & 223,344 & 3,039 $\pm$ 55 & 1.36 $\pm$ 0.02 & 601 $\pm$ 25 & 0.27 $\pm$ 0.01 & 0.198 \\  
\isotope[40][]Ar & 231,893 & 4,007 $\pm$ 63 & 1.73 $\pm$ 0.03 & 1,325 $\pm$ 36 & 0.57 $\pm$ 0.02 & 0.331 \\ 
\isotope[84][]Kr & 249,731 & 5,566  $\pm$ 75 & 2.23 $\pm$ 0.03 & 2,435 $\pm$ 49 & 0.98 $\pm$ 0.02 &  0.437 \\
\isotope[132][]Xe & 249,858 & 6,032 $\pm$ 78 & 2.41 $\pm$ 0.03 & 2,915 $\pm$ 54 & 1.17 $\pm$ 0.02 & 0.483 \\
\bottomrule
\end{tabular}
\label{tab:gibuu_stats}
\end{table*}

\noindent In ABLA++, the hypersources and their excitation energies are then used as an input to determine their deexcitation via fission or evaporation of either protons, neutrons or other light particles based on the Weisskopf formalism~\cite{PhysRev.52.295} and finally gammas until a ground state hypernucleus is reached. The code has been used to validate the deexcitation and evaporation of residues~\cite{boudard2013new,mancusi2014extension} following the fission of \isotope[181][]{Ta}{}~\cite{ayyad2014proton} and \isotope[208][]{Pb}{}~\cite{joseluis2014protoninduced,joseluis2015complete,ayyad2015dissipative,PhysRevC.94.061601},  and recently to investigate the hypernuclear formation in nucleon-induced spallation reactions~\cite{rodriguez2022hypernuclei,PhysRevC.98.021602,joseluis2023constraint}. The latter was made possible by implementation of the Bethe-Weizsäcker mass formula and its hypernuclear extension, which allows the calculation of particle separation energies as the difference in binding energy of the initial (hyper-)nucleus and the residual (hyper-)nucleus after a potential evaporation. Besides the excitation energy and baryon content, the momentum, kinetic energy and angular momentum of the hypersource have to be provided. The latter is not directly accessible in GiBUU. It is assumed that low angular momentum states are predominantly populated, so that $J=2$ was set for all deexcitation processes.

\section{Yields}
\label{Yields}

\begin{figure}[H]
    \centering
    \includegraphics[width=.98\linewidth]{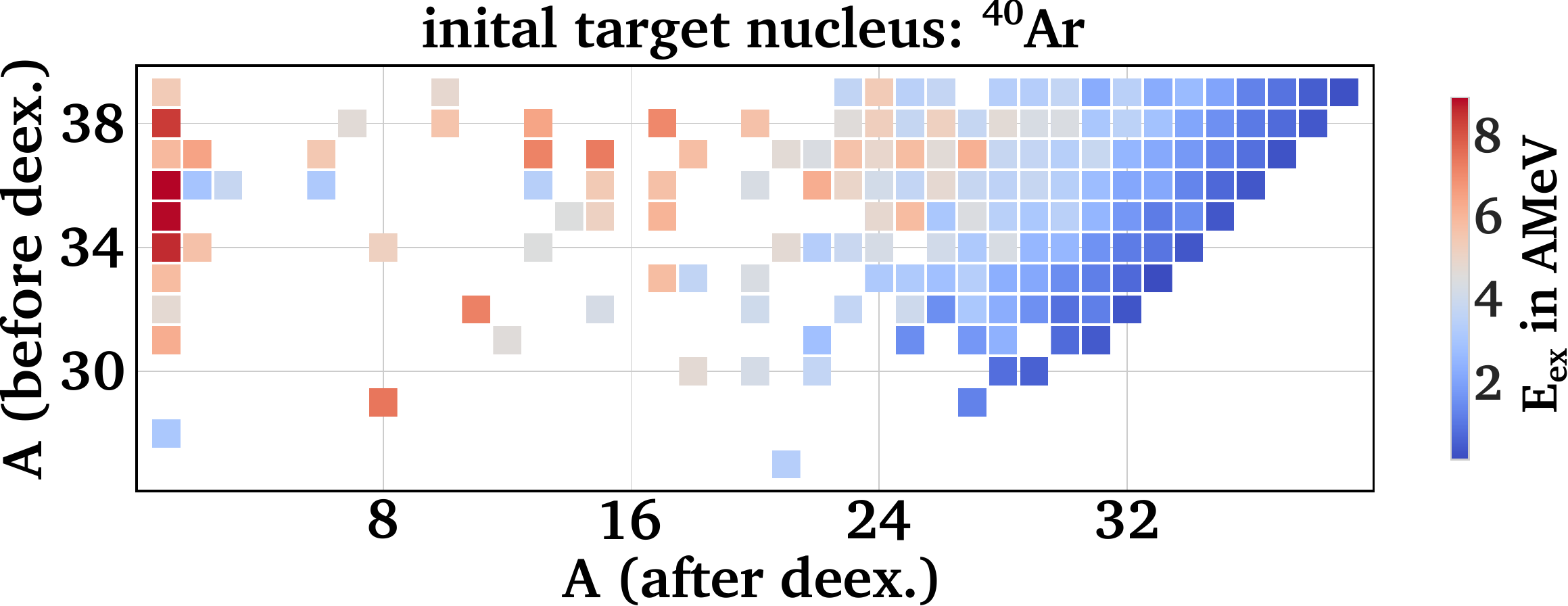}
    \caption{Dependence of the final g.s. hypernuclei on the initial excitation energy of the corresponding hypersource derived from GiBUU for the case of \isotope[40][]{Ar}{}.}
    \label{fig:corr_eex_final_A}
\end{figure}

Based on previous experimental and simulation \linebreak studies on strangeness production in antiproton annihilations, the expected rate of \textLambda\,production is in the order of about 1\,$\%$ to 3\,$\%$ per annihilation, depending on the target nucleus~\cite{cugnon1990strangeness,feng20,epherre1989production,balestra1991strangeness}. 
As the annihilations treated in the following simulations occur on the nuclear surface, it is expected that not all \textLambda s will be captured by the residual nucleus to form an excited hypersource. Thus, to get an overview over the accessible range of hypernuclei after deexcitation, a sufficient number of annihilations has to be considered. \\
For each target nucleus, about 250,000 initial collisions were simulated. Due to the peripheral initialization with large collision impact parameter, an annihilation occurs in about 90 to 95\,\% of all collisions, while otherwise the antiproton simply passes the nucleus without an interaction. The amount of simulated annihilations and production rates of \textLambda\,baryons and hypersources per annihilation are presented in Table \ref{tab:gibuu_stats}. The rate of \textLambda\,baryons is consistent with the previous measurements with \linebreak stopped antiproton beams at LEAR~\cite{bocquet1987delayed,armstrong93}, and between 20\,$\%$ to 50\,$\%$ of the \textLambda\,hyperons can form a hypersource after interacting with the residual nucleus, while the capture probability increases with the number of residual nucleons. To estimate the statistical uncertainty of the yields, we assume that the production of \textLambda\,baryons and hypersources follows a Poisson distribution, as all simulations are performed within the standard parallel-ensemble method over the same time interval~\cite{buss2012transport}. \\ 
Typically, a few protons and neutrons are evaporated by the initial annihilation, leading to a high abundance of hypersources with baryon mass numbers close to the initial nucleus. The excitation energies range from 0 up to 10\,MeV/nucleon with a mean value of about 1.5\,MeV/nucleon for \linebreak \isotope[132][]Xe, 1.9\,MeV/nucleon for \isotope[84][]Kr, 2.4\,MeV/nucleon for \isotope[40][]Xe, and 4.2\,MeV/nucleon for \isotope[16][]O. This mean excitation energy per nucleon scales approximately inversely with the square root of the ratio of baryon numbers, leading to a total excitation energy that scales up with the square root of the baryon numbers. The absolute values for the excitation energies are comparable with other production approaches for hypernuclei, such as relativistic light ion collisions~\cite{sun2018production}. As the typical nucleon separation energy is about 10-15\,MeV and the alpha particle removal energy is about 35\,MeV, most hypersources will deexcite via the emission of a few alpha particles and several nucleons, leading to overall lighter ground state hypernuclei. Figure \ref{fig:corr_eex_final_A} depicts the dependence of the hypersource mass number before the deexcitation and the final ground state hypernucleus mass number depending on the initial excitation energy per nucleon for the case of \isotope[40][]{Ar}. While low excitation energies up to about 3\,MeV/nucleon lead to the production of heavy g.s. hypernuclei, the \textLambda\,emission channel becomes accessible for higher excitation energies, leading to free \textLambda\,hyperons ($A=1$) in the final state of the deexcitation. \\ 
\begin{figure*}[hptb]
  \label{}
    \centering
    \includegraphics[width=0.90\linewidth]{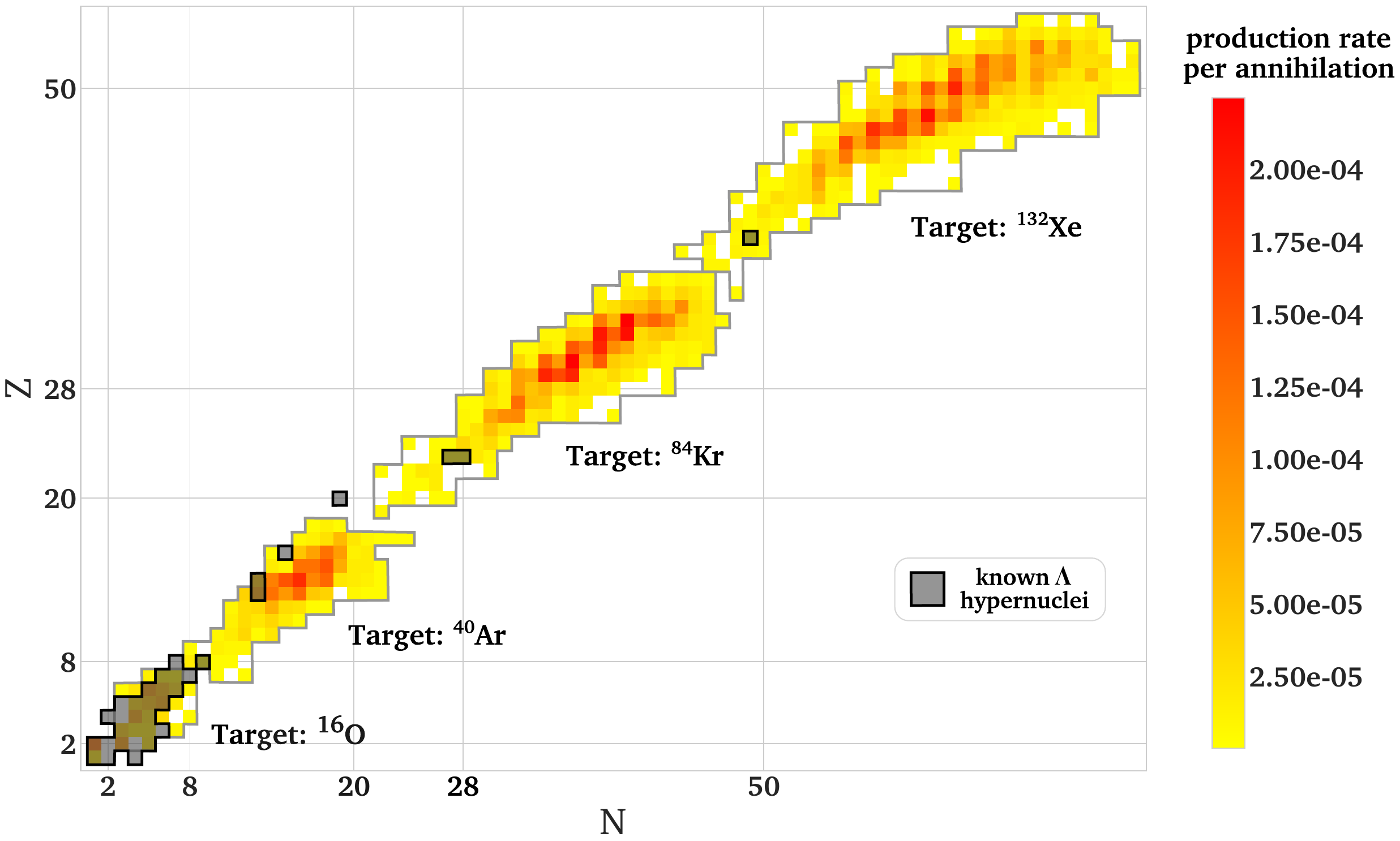} 
\caption{Yields of different hypernuclear isotopes per annihilation on the respective target nucleus. The landscape of already investigated hypernuclei is shaded in gray~\cite{chartmainz}.} 
\label{fig:deexcited_gs_hypernuclei}
\end{figure*}
\begin{table*}[hptb]
\centering
\caption{Most abundant g.s. hypernuclei after deexcitation in ABLA++ for the four reference nuclei.} 
\begin{tabular}{ c  c  c  c  c  c  c  c }
\multicolumn{2}{c}{\isotope[16][]O} & \multicolumn{2}{c}{\isotope[40][]Ar} & \multicolumn{2}{c}{\isotope[84][]Kr} & \multicolumn{2}{c}{\isotope[132][]Xe} \\
\midrule
nucleus&rate in $10^{-5}$&nucleus&rate in $10^{-5}$&nucleus&rate in $10^{-5}$&nucleus&rate in $10^{-5}$\\
\midrule
\isotope[10][\text{\textLambda}]B & 9 $\pm$ 2 & \isotope[29][\text{\textLambda}]Al & 19 $\pm$ 3  & \isotope[65][\text{\textLambda}]Cu & 28 $\pm$ 3  & \isotope[109][\text{\textLambda}]Ag & 21 $\pm$ 3  \\
\isotope[7][\text{\textLambda}]Li & 8 $\pm$ 2 & \isotope[32][\text{\textLambda}]Si & 15 $\pm$ 3  & \isotope[72][\text{\textLambda}]Ge & 24 $\pm$ 3  & \isotope[113][\text{\textLambda}]In & 19 $\pm$ 3  \\
\isotope[10][\text{\textLambda}]Be & 7 $\pm$ 2 & \isotope[28][\text{\textLambda}]Al & 15 $\pm$ 3  & \isotope[71][\text{\textLambda}]Ga & 21 $\pm$ 3  & \isotope[104][\text{\textLambda}]Pd & 18 $\pm$ 3  \\
\isotope[11][\text{\textLambda}]B & 6 $\pm$ 2 & \isotope[31][\text{\textLambda}]Al & 13 $\pm$ 2  & \isotope[69][\text{\textLambda}]Ga & 20 $\pm$ 3  & \isotope[106][\text{\textLambda}]Pd & 16 $\pm$ 3  \\
\isotope[9][\text{\textLambda}]Be & 5 $\pm$ 2 & \isotope[28][\text{\textLambda}]Mg & 13 $\pm$ 2  & \isotope[62][\text{\textLambda}]Ni & 19 $\pm$ 3  & \isotope[107][\text{\textLambda}]Ag & 16 $\pm$ 3  \\
\bottomrule
\end{tabular}
\label{tab:most_abundant_HN}
\end{table*}
\noindent Figure \ref{fig:deexcited_gs_hypernuclei} summarizes the yield of ground state hypernuclei after the deexcitation simulation in \linebreak ABLA++ relative to an annihilation process. For each initial nucleus, a predominant region of production arises with production rates of 1 to 2 per mille for several hyperisotopes indicated by red-shaded colors, indicating that a wide range of so-far undiscovered hypernuclei becomes accessible. \\
Similarly to the total excitation energy, the difference of the initial nucleus baryon number and the high production g.s. hypernuclei baryon number rises with the square root of the initial baryon numbers. In Table \ref{tab:most_abundant_HN} the five most abundant g.s. hypernuclei and their simulated production rates are summarized for each target nucleus. Combining the yields from all four initial nuclei, over 200 new hypernuclei in the medium-mass region of the hypernuclear chart are accessible. We underline here that an experimental method to tag and identify produced hypernuclei still needs to be implemented. While this work focuses on the prediction of production yields, effort towards an experimental implementation are ongoing.  \\
The rate of production for a hypernucleus of interest reaches up to a few $10^{-4}$ per annihilation. Consequently, at least tens of millions antiproton-nucleus annihilations have to be considered to provide sufficient statistics for a significant measurement. Such conditions can be reached at the extra-low energy antiproton (ELENA) ring of CERN~\cite{maury2014elena}, where up to 10$^7$ antiprotons/bunch are provided every 120 seconds, so that a measurement of a specific hypernucleus could be performed within a few hours for the considered stable reference nuclei. The properties of the produced hypernuclei, \textit{e.g.} their ground state mass, can then be accessed based on their mesonic weak decay via an invariant mass measurement. While the light hypernuclei produced from oxygen decay predominantly via meson emission, the non-mesonic decay \textLambda N $\rightarrow $ NN is dominant for the heavier hypernuclei originating from argon, krypton and xenon~\cite{sato2005mesonic,vidana2018hyperons}. \\
The authors have been made aware of a recent first implementation of antiproton annihilations at rest and in-flight within the INCL nuclear cascade transport code, integrated inside GEANT4~\cite{Geant4antiproton}. When available, a comparison of this model's predictions to the present work for hypernuclei production would be of interest.

\begin{table*}[hptb]
\centering
\caption{Statistics of the strangeness production from antiproton annihilations in GiBUU assuming only strangeness exchange (SE) or non-strange meson interactions (NSMI).}
\begin{tabular}{ c  c  c  c  c  c  c }
\toprule
Nucleus & open channel & annihilations & \textLambda s  & \textLambda\,rate in $\%$ & HSs & HS rate in $\%$\\
\midrule
\multirow{2}{*}{\isotope[40][]Ar} & SE & 9,230 & 180 $\pm$ 13 & 1.95  $\pm$ 0.15 & 66  $\pm$ 8 & 0.71 $\pm$ 0.09 \\
& NSMI & 9,217 & 50 $\pm$ 7 & 0.54  $\pm$ 0.08 & 14 $\pm$ 4  & 0.15 $\pm$ 0.04  \\ 
\midrule
\multirow{2}{*}{\isotope[84][]Kr} & SE & 9,994 & 230 $\pm$ 15 & 2.30 $\pm$ 0.15 & 102 $\pm$ 10 & 1.02 $\pm$ 0.10 \\
& NSMI & 9,995 & 89  $\pm$ 9 & 0.89 $\pm$ 0.09 & 22 $\pm$ 5 & 0.22 $\pm$ 0.05 \\ 
\bottomrule
\end{tabular}
\label{tab:production_channels}
\end{table*}

\subsection{Strangeness Production Mechanism}
\label{Production}

To investigate how the \textLambda\,baryons are produced subsequently to the initial annihilation in GiBUU, the cross sections for strangeness exchange reactions of nucleons with kaons and the strangeness production by pions as well as \textomega \, and \texteta \, mesons have been set to zero separately. \\
The two target nuclei \isotope[40][]Ar and \isotope[84][]Kr have been considered to examine the ratio of the production channels for different masses of target nuclei. For all simulations, the \textLambda\,\,and \textSigma\,\,production channels are considered together due to the strong coupling between the two baryons. The simulations were performed with similar initial conditions. The resulting \textLambda\,baryon and hypersource production rates are summarized in Table \ref{tab:production_channels}.\\
For both nuclei, the strangeness exchange with primary kaons is the dominant \textLambda\,and hypersource production mechanism, contributing to about 82\,\%, while the remaining 18\,\% are produced via nucleon non-strange meson interactions. The \textLambda\,baryons produced in a strangeness exchange reaction have a probability of about 40 to 50\,\% to be created in a low-momentum state which allows the attribution to a hypersource, while only about 25\,\% of the \textLambda\,produced by pion interactions can be captured by the residual nucleus.

\subsection{Isospin dependence}
\label{Isospin}

\begin{figure*}[hptb]
    \centering
    \includegraphics[width=.999\linewidth]{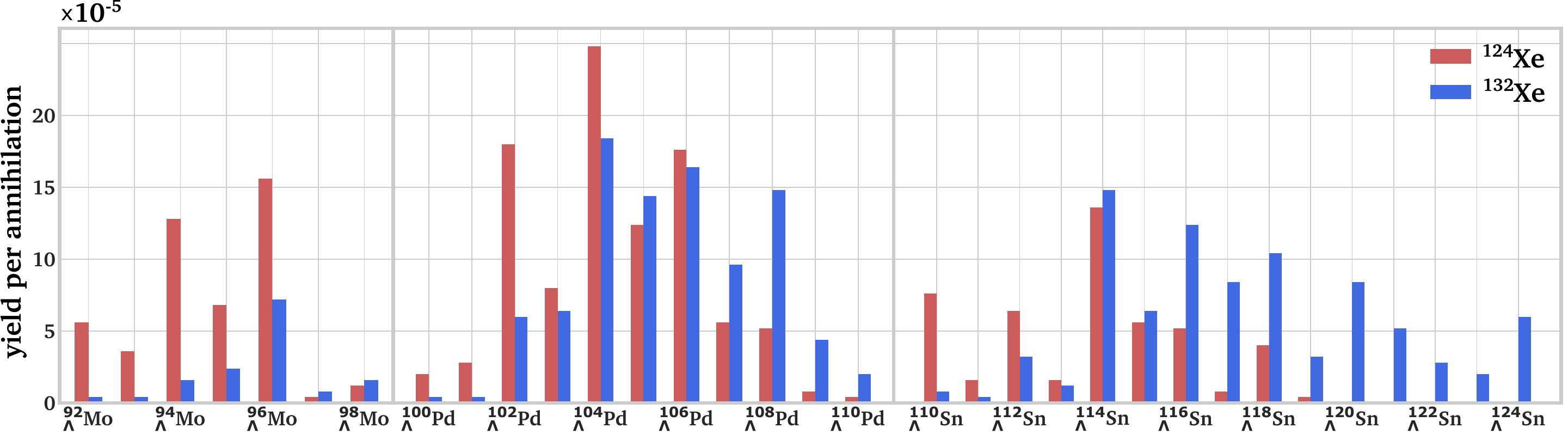}
    \caption{Impact of the initial target nucleus on the yields along different hyperisotopic chains. The neutron-deficient \isotope[124][]Xe favors the production of lighter and more neutron-deficient isotopes in the \isotope[][\text{\textLambda}]Mo and \isotope[][\text{\textLambda}]Pd chains , while \isotope[132][]Xe provides high yields in neutron-rich and heavy isotopes along the \isotope[][\text{\textLambda}]Pd and \isotope[][\text{\textLambda}]Sn chains.}
    \label{fig:isochains_yields}
\end{figure*}
For the simulations presented above, the most abundant isotope of each reference target nucleus was taken into account. However, there are other stable isotopes of, {\itshape e.g.}, argon and xenon, which can be considered for the production of more neutron-deficient hypernuclei due to a lower neutron number in the initial target nucleus. In the following, simulations are presented for \isotope[36][]{Ar} and \isotope[124][]Xe and the hypernuclear yields are compared to the results of \isotope[40][]Ar and \isotope[132][]Xe, respectively. While the overall production rate of hypernuclei remains similar, the yields of individual hyperisotopes shift according to the initial neutron number. Figure \ref{fig:isochains_yields} summarizes the yields of different ground state hypernuclei along different isotopic chains based on annihilations on \isotope[124][]Xe and \isotope[132][]Xe. While the heavier and neutron-rich target isotope favors the production of high $Z$ and high $N$ hypernuclei along the \isotope[][\text{\textLambda}]Sn chain, the neutron-deficient target favors the production of more neutron- and proton-deficient hypernuclei along the \isotope[][\text{\textLambda}]Mo chain. A similar trend is also observed for the case of argon isotopes, with higher yields of \isotope[][\text{\textLambda}]Si isotopes for \isotope[40][]{Ar}{} and a strong abundance of the \isotope[][\text{\textLambda}]Mg chain for \isotope[36][]{Ar}{}. To cover an as wide range of the hypernuclear chart as possible, it is thus useful to consider the full range of stable target isotopes.

\section{Conclusion}
\label{Conclusion}

The production of single \textLambda-hypernuclei from the peripheral annihilation of a captured antiproton following the formation of antiprotonic atoms has been investigated systematically. We report on \linebreak Monte-Carlo simulations within the GiBUU transport framework, while the deexcitation of hyperfragments is performed with the ABLA++ code. The sites of annihilations are derived by fitting to the overlap of the nuclear density with analytical radial antiprotonic orbital wavefunctions from QED. Following the initial annihilation, the produced mesons interact with the residual nucleus and produce \textLambda\,hyperons in about 1\,$\%$ to 3\,$\%$ of annihilations, consistent with previous measurements. The \textLambda\,hyperons are predominantly produced via direct strangeness exchange with primary kaons produced from annihilations in $\sim$80\,\% of the cases, while pion-induced strangeness production contri-butes to the remaining $\sim$20\,\%. The produced \textLambda\, \linebreak hyperons may then be captured by the residual nucleons to form an excited hypersource based on phase-space coalescence, where the hypersource rate per \textLambda\,increases with the initial nucleus mass number from about 20\,$\%$ for \isotope[16][]O to about 50\,$\%$ for \isotope[132][]Xe. The deexcitation via evaporation of the excited hypersources determines the yields and isotopic composition of ground state hypernuclei. The production rates of specific g.s. hypernuclei reach up to a few $10^{-4}$ per annihilation. While the statistical uncertainties of the presented results for the production rates in GiBUU are guided by the number of initial annihilations simulated, the uncertainties of the yields extracted from ABLA++ originate predominantly from the uncertainty of the assumed hypernuclear binding energies and the corresponding excitation energies. While this uncertainty might shift the yields of individual hyperisotopes, the overall conclusion in terms of range of accessible hypernuclei based on antiprotonic annihilations does remain. \\
The reported results based on peripheral annihilations of antiprotons on nuclei indicate that a wide range of ground state hypernuclei can be populated with competitive yields. By considering different stable target isotopes, long hyperisotopic chains can be populated. Additional information about hypernuclear properties and the underlying \textLambda N and \textLambda NN interactions could be gained by investigating their weak decay. This work, along with the software framework used, constitutes a first step towards exploring the feasibility of a dedicated experiment at the ELENA accelerator at CERN. \\ \\


\textbf{Acknowledgements} This work was supported by the European Research Council through the ERC grant PUMA-726276 and the Alexander-von-Hum-boldt foundation. We thank Meytal Duer for discussions on the content and the proof-reading of the manuscript as well as Jean-Christophe David for his comments. J.L. R.-S. is grateful for the support provided by the 'Ramón y Cajal' program under Grant No. RYC2021-031989-I.

\printbibliography

\newpage
\onecolumn
\appendix

{\fontsize{9}{10}\selectfont

\section{Annihilation Branching Ratios}

\begin{table*}[htb]
    \caption*{Branching ratios of final states for antiproton-proton annihilations (left) and antiproton-neutron annihilations (right). In case of non-strange final states, only channels with at least 2\,$\%$ probability are depicted, and for kaonic final states with at least 0.1\,\%.}
    \label{tab:branching_ratios}
    
    \begin{tabular}[t]{lc}
                \multicolumn{2}{c}{Antiproton - Proton} \\
                \toprule
                
                Final State & Probability in $\%$ \\
                \midrule 
$	\rho^{+}	\rho^{-}					$	&	3.37	\\
$	\pi^{+}	\pi^{-}	\pi^{0}				$	&	2.34	\\
$	\pi^{+}	\pi^{-}	\rho^{0}				$	&	2.02	\\
$	\pi^{+}	\pi^{0}	\rho^{-}				$	&	2.02	\\
$	\pi^{-}	\pi^{0}	\rho^{+}				$	&	2.02	\\
$	\pi^{+}	\pi^{-}	\omega				$	&	3.03	\\
$	\pi^{+}	\pi^{-}	\pi^{0}	\omega			$	&	2.84	\\
$	\pi^{+}	\pi^{+}	\pi^{-}	\pi^{-}			$	&	2.74	\\
$	\pi^{+}	\pi^{-}	\pi^{0}	\pi^{0}			$	&	3.89	\\
$	\pi^{+}	\pi^{+}	\pi^{-}	\rho^{-}			$	&	2.58	\\
$	\pi^{+}	\pi^{-}	\pi^{-}	\rho^{+}			$	&	2.58	\\
$	\pi^{+}	\pi^{-}	\pi^{0}	\rho^{0}			$	&	6.29	\\
$	\pi^{+}	\pi^{0}	\pi^{0}	\rho^{-}			$	&	5.05	\\
$	\pi^{-}	\pi^{0}	\pi^{0}	\rho^{+}			$	&	5.05	\\
$	\pi^{+}	\pi^{+}	\pi^{-}	\pi^{-}	\pi^{0}		$	&	2.61	\\
$	\pi^{+}	\pi^{-}	\pi^{0}	\pi^{0}	\omega		$	&	2.58	\\
$	\pi^{+}	\pi^{+}	\pi^{+}	\pi^{-}	\pi^{-}	\pi^{-}	$	&	2.83	\\
$	\pi^{+}	\pi^{+}	\pi^{-}	\pi^{-}	\pi^{0}	\pi^{0}	$	&	9.76	\\
$	\pi^{+}	\pi^{-}	\pi^{0}	\pi^{0}	\pi^{0}	\pi^{0}	$	&	2.68	\\
$	K^{*+}	K^{*-}					$	&	0.225	\\
$	K^{*0}	\Bar{K}^{*0}					$	&	0.225	\\
$	K^{0}	\Bar{K}^{0}	\pi^{0}				$	&	0.146	\\
$	K^{+}	K^{-}	\pi^{0}				$	&	0.146	\\
$	K^{0}	K^{-}	\pi^{+}				$	&	0.142	\\
$	\Bar{K}^{0}	K^{+}	\pi^{-}				$	&	0.142	\\
$	K^{0}	\Bar{K}^{0}	\omega				$	&	0.232	\\
$	K^{+}	K^{-}	\omega				$	&	0.232	\\
$	K^{0}	\Bar{K}^{0}	\rho^{0}				$	&	0.202	\\
$	K^{+}	K^{-}	\rho^{0}				$	&	0.202	\\
$	K^{0}	K^{-}	\rho^{+}				$	&	0.234	\\
$	\Bar{K}^{0}	K^{+}	\rho^{-}				$	&	0.234	\\
$	K^{*+}	\Bar{K}^{0}	\pi^{-}				$	&	0.23	\\
$	K^{*-}	K^{0}	\pi^{+}				$	&	0.23	\\
                \bottomrule \\
    \end{tabular}
    \qquad \qquad
    \begin{tabular}[t]{lc}

                \multicolumn{2}{c}{Antiproton - Neutron} \\
                \toprule
                
                Final State & Probability in $\%$ \\
                \midrule
$	\rho^{-}	\rho^{0}					$	&	3.51	\\
$	\rho^{-}	\eta					$	&	2.27	\\
$	\rho^{-}	\omega					$	&	3.51	\\
$	\pi^{+}	\pi^{-}	\pi^{-}				$	&	2.86	\\
$	\pi^{+}	\pi^{-}	\rho^{-}				$	&	3.62	\\
$	\pi^{-}	\pi^{0}	\rho^{0}				$	&	5.61	\\
$	\pi^{0}	\pi^{0}	\rho^{-}				$	&	3.51	\\
$	\pi^{-}	\rho^{+}	\rho^{-}				$	&	2.09	\\
$	\pi^{-}	\pi^{0}	\omega				$	&	5.05	\\
$	\pi^{+}	\pi^{-}	\pi^{-}	\omega			$	&	10.52	\\
$	\pi^{-}	\pi^{0}	\pi^{0}	\omega			$	&	7.01	\\
$	\pi^{+}	\pi^{-}	\pi^{-}	\pi^{0}			$	&	5.51	\\
$	\pi^{+}	\pi^{-}	\pi^{-}	\pi^{0}	\pi^{0}		$	&	2.72	\\
$	\pi^{+}	\pi^{+}	\pi^{-}	\pi^{-}	\pi^{-}	\pi^{0}	$	&	8.33	\\
$	\pi^{+}	\pi^{-}	\pi^{-}	\pi^{0}	\pi^{0}	\pi^{0}	$	&	6.67	\\
$	K^{0}	K^{-}					$	&	0.147	\\
$	K^{*-}	K^{*0}					$	&	0.184	\\
$	K^{0}	K^{-}	\pi^{0}				$	&	0.316	\\
$	K^{0}	\Bar{K}^{0}	\pi^{-}				$	&	0.432	\\
$	K^{+}	K^{-}	\pi^{-}				$	&	0.513	\\
$	K^{0}	K^{-}	\omega				$	&	0.35	\\
$	K^{0}	K^{-}	\rho^{0}				$	&	0.15	\\
$	K^{0}	\Bar{K}^{0}	\rho^{-}				$	&	0.77	\\
$	K^{+}	K^{-}	\rho^{-}				$	&	0.77	\\
$	K^{*-}	K^{0}	\pi^{0}				$	&	0.245	\\
$	K^{*0}	K^{-}	\pi^{0}				$	&	0.245	\\
$	K^{*0}	\Bar{K}^{0}	\pi^{-}				$	&	0.13	\\
$	\Bar{K}^{*0} K^{0}	\pi^{-}				$	&	0.13	\\
$	K^{*+}	K^{-}	\pi^{-}				$	&	0.154	\\
$	K^{*-}	K^{+}	\pi^{-}				$	&	0.154	\\

                \bottomrule \\
    \end{tabular}
  
\end{table*}

}
 
\end{document}